\documentclass{elsart}
\usepackage{epsfig}
\usepackage{amssymb}
\usepackage{amsmath}
\newcommand{\affuni}[2]{Dipartimento di Fisica dell'Universit\`a #1, #2, Italy.}
\newcommand{\affinfn}[2]{INFN Sezione di #1, #2, Italy.}

\def\raE     {\rightarrow}
\def \daf{DA$\Phi$NE}
\def \ff{${\it \phi}$-factory}
\def \epem{$e^+e^-$~}

\def \Kp3pi  {${\rm K^+} \rightarrow \pi^+\pi^-\pi^+$~}
\def \Kp3pig {${\rm K^+} \rightarrow \pi^+\pi^-\pi^+(\gamma)$~}
\def \Kpipi  {${\rm K^{\pm}} \rightarrow \pi^{\pm}\pi^0(\gamma)$~}
\def \Kmunu  {${\rm K^{\pm}} \rightarrow \mu^{\pm}\nu(\gamma)$~}
\usepackage{amssymb}
\usepackage{lineno}

\journal{Physics Letters B}

\begin{document}

\begin{frontmatter}



\title{ Measurement of the absolute branching ratio of the $K^+ \raE  
     \pi^+\pi^-\pi^+(\gamma)$  decay with the KLOE detector }


%%%%\author{}
%%%%\address{}
\collab{The KLOE/KLOE-2 Collaboration}
%\author[Roma2,INFNRoma2]{F.~Archilli},
\author[Frascati]{D.~Babusci},
%\author[Roma2,INFNRoma2]{D.~Badoni},
\author[Cracow]{I.~Balwierz-Pytko},
\author[Frascati]{G.~Bencivenni},
%\author[Roma1,INFNRoma1]{C.~Bini},
\author[Frascati]{C.~Bloise},
%\author[INFNRoma1]{V.~Bocci},
\author[Frascati]{F.~Bossi},
\author[INFNRoma3]{P.~Branchini},
\author[Roma3,INFNRoma3]{A.~Budano},
%\author[Moscow]{S.~A.~Bulychjev},
\author[Uppsala]{L.~Caldeira~Balkest\aa hl},
%\author[Frascati]{P.~Campana},
%\author[Frascati]{G.~Capon},
\author[Roma3,INFNRoma3]{F.~Ceradini},
\author[Frascati]{P.~Ciambrone},
\author[Messina,INFNCatania]{F.~Curciarello},
\author[Cracow]{E.~Czerwi\'nski},
\author[Frascati]{E.~Dan\`e},
\author[Messina,INFNCatania]{V.~De~Leo},
\author[Frascati]{E.~De~Lucia},
\author[INFNBari]{G.~De~Robertis},
%\author[Roma1,INFNRoma1]{A.~De~Santis},
\author[Frascati]{A.~De~Santis},
%\author[Roma1,INFNRoma1]{G.~De~Zorzi},
\author[Frascati]{P.~De~Simone\corauthref{cor}},
\author[Roma3,INFNRoma3]{A.~Di~Cicco},
\author[Roma1,INFNRoma1]{A.~Di~Domenico},
%\author[Napoli,INFNNapoli]{C.~Di~Donato},
\author[INFNRoma2]{R.~Di~Salvo},
%\author[Roma3,INFNRoma3]{B.~Di~Micco},
\author[Frascati]{D.~Domenici},
%\author[Frascati]{A.~D'Uffizi},
\author[Bari,INFNBari]{O.~Erriquez},
\author[Bari,INFNBari]{G.~Fanizzi},
\author[Roma2,INFNRoma2]{A.~Fantini},
\author[Frascati]{G.~Felici},
\author[ENEACasaccia,INFNRoma1]{S.~Fiore},
\author[Roma1,INFNRoma1]{P.~Franzini},
\author[Cracow]{A.~Gajos},
\author[Roma1,INFNRoma1]{P.~Gauzzi},
\author[Messina,INFNCatania]{G.~Giardina},
\author[Frascati]{S.~Giovannella},
%\author[Roma2,INFNRoma2]{F.~Gonnella},
\author[INFNRoma3]{E.~Graziani},
\author[Frascati]{F.~Happacher},
\author[Uppsala]{L.~Heijkenskj\"old}
\author[Uppsala]{B.~H\"oistad},
%\author[Frascati]{L.~Iafolla},
%\author[Energetica,Frascati]{E.~Iarocci},
%\author[Uppsala]{M.~Jacewicz},
\author[Uppsala]{T.~Johansson},
%\author[Cracow]{K.~Kacprzak},
\author[Cracow]{D.~Kami\'nska},
\author[Cracow]{W.~Krzemien},
%\author[Warsaw]{A.~Kowalewska},
%\author[Moscow]{V.~Kulikov},
\author[Uppsala]{A.~Kupsc},
\author[Frascati,StonyBrook]{J.~Lee-Franzini},
%\author[Frascati]{B.~Leverington},
\author[INFNBari]{F.~Loddo},
\author[Roma3,INFNRoma3]{S.~Loffredo},
\author[Messina,INFNCatania,CentroCatania]{G.~Mandaglio},
\author[Moscow]{M.~Martemianov},
\author[Frascati,Marconi]{M.~Martini},
\author[Roma2,INFNRoma2]{M.~Mascolo},
%\author[Moscow]{M.~Matsyuk},
\author[Roma2,INFNRoma2]{R.~Messi},
\author[Frascati]{S.~Miscetti},
\author[Frascati]{G.~Morello},
\author[INFNRoma2]{D.~Moricciani},
\author[Cracow]{P.~Moskal},
%\author[INFNRoma3,LIP]{F.~Nguyen},
%\author[Frascati]{L.~Quintieri},
\author[Frascati]{A.~Palladino},
\author[INFNRoma3]{A.~Passeri},
\author[Energetica,Frascati]{V.~Patera},
\author[Roma3,INFNRoma3]{I.~Prado~Longhi},
\author[INFNBari]{A.~Ranieri},
%\author[Mainz]{C.~F.~Redmer},
\author[Frascati]{P.~Santangelo},
\author[Frascati]{I.~Sarra},
\author[Calabria,INFNCalabria]{M.~Schioppa},
\author[Frascati]{B.~Sciascia},
%\author[Energetica,Frascati]{A.~Sciubba},
\author[Cracow]{M.~Silarski},
%\author[Calabria,INFNCalabria]{S.~Stucci},
%\author[Roma3,INFNRoma3]{C.~Taccini},
\author[INFNRoma3]{L.~Tortora},
\author[Frascati]{G.~Venanzoni},
%\author[Frascati,CERN]{R.~Versaci},
\author[Warsaw]{W.~Wi\'slicki},
\author[Uppsala]{M.~Wolke}
%\author[Cracow]{J.~Zdebik}

%%%%%%%%%%%%%%%%%%%%%%%%%%%%%%%%%%%%%%%%%%%%%%%%%%%%%%%%%%%%%%%%%%%%%%%%%%%%%%%%%%%%%%%%%%%%%%%%%%
\corauth[cor]{Corresponding author. {\it Email address:}{\tt patrizia.desimone@lnf.infn.it}}

\address[Bari]{\affuni{di Bari}{Bari}}
\address[INFNBari]{\affinfn{Bari}{Bari}}
\address[CentroCatania]{Centro Siciliano di Fisica Nucleare e Struttura della Materia, Catania, Italy.}
\address[INFNCatania]{\affinfn{Catania}{Catania}}
\address[Calabria]{\affuni{della Calabria}{Cosenza}}
\address[INFNCalabria]{INFN Gruppo collegato di Cosenza, Cosenza, Italy.}
\address[Cracow]{Institute of Physics, Jagiellonian University, Cracow, Poland.}
\address[Frascati]{Laboratori Nazionali di Frascati dell'INFN, Frascati, Italy.}
%\address[Messina]{\affuni{di Messina}{Messina}}
%\address[Mainz]{Institut f\"ur Kernphysik, 
%Johannes Gutenberg Universit\"at Mainz, Germany.}
\address[Messina]{Dipartimento di Fisica e Scienze della Terra dell'Universit\`a di Messina, Messina, Italy.}\address[Moscow]{Institute for Theoretical and Experimental Physics (ITEP), Moscow, Russia.}
%\address[Napoli]{\affuni{``Federico II''}{Napoli}}
%\address[INFNNapoli]{\affinfn{Napoli}{Napoli}}
\address[Energetica]{Dipartimento di Scienze di Base ed Applicate per l'Ingegneria dell'Universit\`a 
``Sapienza'', Roma, Italy.}
\address[Marconi]{Dipartimento di Scienze e Tecnologie applicate, Universit\`a ``Guglielmo Marconi", Roma, Italy.}
\address[Roma1]{\affuni{``Sapienza''}{Roma}}
\address[INFNRoma1]{\affinfn{Roma}{Roma}}
\address[Roma2]{\affuni{``Tor Vergata''}{Roma}}
\address[INFNRoma2]{\affinfn{Roma Tor Vergata}{Roma}}
\address[Roma3]{Dipartimento di Matematica e Fisica dell'Universit\`a 
``Roma Tre'', Roma, Italy.}
%\address[Roma3]{\affuni{``Roma Tre''}{Roma}}
\address[INFNRoma3]{\affinfn{Roma Tre}{Roma}}
\address[ENEACasaccia]{ENEA UTTMAT-IRR, Casaccia R.C., Roma, Italy}
\address[StonyBrook]{Physics Department, State University of New 
York at Stony Brook, USA.}
\address[Uppsala]{Department of Physics and Astronomy, Uppsala University, Uppsala, Sweden.}
\address[Warsaw]{National Centre for Nuclear Research, Warsaw, Poland.}
%\address[CERN]{Present Address: CERN, CH-1211 Geneva 23, Switzerland.}
%\address[LIP]{Present Address: Laborat\'orio de Instrumenta\c{c}\~{a}o e F\'isica Experimental de Part\'iculas,
%Lisbon, Portugal.}
%%%%%%%%%%%%%%%%%%%%%%%%%%%%%%%%%%%%%%%%%%%%%%%%%%%%%%%%%%%%%%%%%%%%%%%%%%%%%%%%%%%%%%%%%%%%%%%%%

\begin{abstract} 
The absolute branching ratio of the $K^+ \raE \pi^+\pi^-\pi^+(\gamma)$ decay,
inclusive of final-state radiation, has been measured using
$\sim$17 million tagged $K^+$  mesons collected with the
KLOE detector at ~\daf, the Frascati ~\ff. 
The result is:
\[
BR(K^+ \raE \pi^+\pi^-\pi^+(\gamma)) = 0.05565 \pm 0.00031_{stat} \pm 0.00025_{syst}
\]

\noindent
%with an overall fractional accuracy of 0.72$\%$, 
a factor $\simeq$ 5 more precise with respect
to the previous result. This work completes the program of precision measurements
of the dominant kaon branching ratios at KLOE.
\end{abstract}


\end{frontmatter}


%===========================
\section{Introduction}
%===========================

The measurement of the branching ratio (BR) of \Kp3pig  ~decay completes the KLOE program of
precision measurements of the dominant kaon branching ratios, fully inclusive of radiation effects.
We have already published an evaluation, from a fit to the KLOE measurements 
of the charged kaon lifetime \cite{Kpmlife}, and BRs, 
\cite{Kmunu}, \cite{KpmSemil}, \cite{Kpmpch2p0}, \cite{KLOE_K3piFit}, constraining the BR sum to unity :
BR($K^{\pm}\rightarrow \pi^{\pm}\pi^+ \pi^-)=(5.68 \pm 0.22)\%$ ~\cite{KLOE_K3piFit}. 
The most recent BR($K^{\pm}\rightarrow \pi^{\pm}\pi^+ \pi^-)$ measurement, 
based on 2330 events from a sample of $\sim$ 10$^5$ kaon decays,
dates back to 1972 and  gives no information on the radiation cut-off : 
BR($K^{\pm}\rightarrow \pi^{\pm}\pi^+ \pi^-)=(5.56 \pm 0.20)\%$  \cite{chiang}.
The PDG value,  BR($K^{\pm}\rightarrow \pi^{\pm}\pi^+ \pi^-)=(5.59\pm 0.04)\%$ \cite{PDG},
is obtained from a global fit that does not use any of the available BR($K^{\pm}\rightarrow \pi^{\pm}\pi^+ \pi^-)$ measurements
but the rate measurement $\Gamma(\pi^+\pi^+\pi^-) = (4.511 \pm 0.024)\times 10^6$~s$^{-1}$ 
published in 1970  \cite{Ford}.
%without information about the treatment of final-state radiation
Furthermore the BR($K^{\pm}\rightarrow \pi^{\pm}\pi^+ \pi^-)$ value enters in the
evaluation of the difference $a_0 - a_2$ between the $I = 0$ and $I = 2$ $S$-wave $\pi\pi$
scattering lengths \cite{NA48-1} \cite{NA48-2}; this will be discussed in section 5.

\noindent
In the following we report the measurement of the absolute branching ratio  
BR$(K^+ \raE \pi^+\pi^-\pi^+(\gamma))$ performed with the KLOE detector 
using data corresponding to an
integrated luminosity $\int \cal{L}$ dt $\simeq$ 174 pb$^{-1}$ collected at 
\daf, the Frascati ~\ff \cite{DAFNE}. \daf~ is an ~\epem collider  
operated at the energy of 1020 MeV, the mass of the $\phi$-meson.
%working at a center of mass energy of 
%$\sqrt s \simeq m_{\phi} \sim 1020$ MeV. The $\phi$
%production cross section is $\sim 3 \mu \mbox{b}$.
The beams collide at the interaction point (IP) with a crossing angle 
$\theta_x \simeq 25$ mrad 
\footnote{We use left-handed coordinates system with the $z$-axis defined as the
bisectrix of the ~\epem beams and the $y$-axis vertical.}, 
producing $\phi$-mesons 
%therefore the $\phi$'s are produced 
with a small momentum of $\sim 12.5$ MeV in the horizontal plane.
The  $\phi$-mesons decay in
anti-collinear and monochromatic neutral ($34 \%$) and 
charged ($49 \%$) kaon pairs.
The unique feature of a ~\ff~ is the tagging:
detection of a $K^{\pm}$ (the tagging kaon) tags the presence of a
$K^{\mp}$ (the tagged kaon) with known momentum and direction.
The availability of tagged kaons enables the   
precision measurement of absolute BRs providing the normalization sample.
The decay products of the $K^+K^-$ pair define two spatially
well separated regions called in the following the tag and the
signal hemispheres.

%==============================================================================
\section{The KLOE detector}
%==============================================================================

The KLOE detector consists of a large cylindrical drift chamber (DC)~\cite{DCH},
surrounded by a lead scintillating fiber electromagnetic calorimeter
(EMC)~\cite{EMC} both immersed in an axial 0.52~T magnetic field produced by a
su\-per\-con\-du\-cting coil. At the beams IP
the spherical beam pipe of 10~cm radius is made of a beryllium-aluminum alloy
of 0.5 mm thickness.

\noindent
The DC tracking system has
25 cm internal radius, 4 m diameter and 3.3 m length,
with a total of $\sim 52000$ wires, of which $\sim 12000$ are sense wires
arranged in a stereo geometry.
In order to minimize the multiple scattering and 
$K_L$ regeneration, and to maximize the detection efficiency for low energy photons,
the DC works with a helium-based gas mixture and its walls are made of
light materials, mostly carbon fiber composites. 
Spatial resolutions are $\sigma_{xy} \simeq 150\ \mu$m and 
$\sigma_z \simeq $~2 mm and
the transverse momentum resolution is $\sigma(p^T)/p^T \le 0.4\%$.
%the momentum resolution is $\sigma(p_{\perp})/p_{\perp}\le 0.4\%$.

\noindent
The calorimeter covers 98\% of the solid angle and is
composed by a barrel  and two endcaps.
Particles showering in the lead-scintillator-fiber EMC
stru\-ctu\-re are detected as local energy deposits by clustering signals from
read-out elements. For each impinging particle the calorimeter information consists of energy, 
position of impact point and time of arrival  with accuracies of 
$\sigma_E/E$ = 5.7$\%/\sqrt{E\ {\rm(GeV)}}$ ,
$\sigma_z$ = 1.2 cm/$\sqrt{E\ {\rm(GeV)}}$ , $\sigma_{\phi}$ = 1.2 cm, and 
$\sigma_t = 57\ {\rm ps}/\sqrt{E\ {\rm(GeV)}} \oplus100\ {\rm ps}$.
Energy clusters not associated with reconstructed 
tracks in the DC (neutral clusters) identify neutral particles.
The definition of energy clusters associated with reconstructed tracks is related to
the track-to-cluster association procedure described in Ref. \cite{offKLOE}.

\noindent
The trigger \cite{TRG}
is based on energy deposits in the calorimeter and on 
hit multiplicity in the drift chamber. Only events triggered by the calorimeter 
have been used in the present analysis. 
The trigger system includes a second-level veto for cosmic-ray muons (cosmic-ray veto
or CRV) based on energy deposits in the outermost layers of the
calorimeter and followed by a third-level software trigger. 
%able to identify most of the $\phi$-meson decays. 
A software filter (SF), based on the topology
and multiplicity of energy clusters and drift chamber hits, is applied
to filter out machine background. Both CRV and SF 
%are sources of event rejection.
may be sources of events loss.
Their effect on the BR measurement has been studied on 
%data control samples which do not undergo, respectively 
control data samples acquired respectively without
the CRV and the SF filters.

\noindent
The data sample used for this analysis has been processed and filtered 
with the KLOE standard reconstruction software and the event classification procedure \cite{offKLOE}.
The KLOE monte carlo (MC) simulation package, GEANFI, has been used to produce a
sample equivalent to data, accounting for
the detector status and the machine operation on a run-by-run basis.

%==============================================================================
\section{Analysis strategy }
%==============================================================================

Tagging with  ~\Kmunu ($K_{\mu2}$ tags) and ~\Kpipi ($K_{\pi2}$ tags)
provides two indipendent samples of pure kaons for the signal selection
useful for systematic uncertainties evaluation and cross-checks \cite{KpmSemil}.
These decays are easily 
identified as clear peaks in the distribution of p$^{\ast}_{m_\pi}$, the momentum of the charged secondary
track in the kaon rest frame evaluated using the pion mass
\footnote{The contribution to the p$^{\ast}_{m_\pi}$ distribution from $K_{\mu2}$ 
decays is slightly broadened due to the pion mass hypothesis \cite{KpmSemil}.}.
The selection efficiency of the two tagging normalization samples are similar, about 36 \%.
Then these events are classified as
$\phi \raE K^+K^-$ and archived in dedicated data summary tapes, as described in  Ref. \cite{offKLOE}.
MC studies show that the contamination due to $\phi$-meson decays other than $K^+K^-$ is negligible.

\noindent
To minimize the impact of the trigger efficiency on the signal side,
we choose as normalization sample $K_{\mu2}$ or $K_{\pi2}$ tags
providing the trigger of the event
(self-triggering two-body decays).
After this request the $K_{\mu2}$ sample is reduced by a factor of $\sim$ 33\%, 
while the $K_{\pi2}$ sample by a factor of $\sim$ 43\%.
The residual dependence of the signal sample on the tag selection,
which we refer to as tag bias, has been evaluated for the BR measurement.
Moreover we use  $K^-$ as the tagging kaon ($K_{\mu2}$ or $K_{\pi2}$) and 
$K^+$ as the tagged kaon (signal),
since the nuclear cross section for positive kaons with momenta $\simeq$
100 MeV is lower by a factor of $\sim 10^3$ with
respect to that of negative kaons \cite{Knuclear}. 

\noindent
The track of the tagging kaon is backward extrapolated from its first hit in the DC to the IP. 
We use the momentum of the tagging kaon at the IP, $\bold{p}_{K^-}^{IP}$, and
the momentum of the $\phi$-meson measured run by run with Bhabha scattering events, $\bold{p}_{\phi}$, 
to evaluate the momentum of the tagged kaon at the IP,  
$\bold{p}_{K^+}^{IP} = \bold{p}_{\phi} -  \bold{p}_{K^-}^{IP}$.
Finally we extrapolate $\bold{p}_{K^+}^{IP}$ inside the DC
(signal kaon path).

\noindent
The kaon and the three charged pions from its decay
have low momenta, less than $\sim200$
MeV, and curl up in the KLOE magnetic field; this increases the
probability to have poorly reconstructed tracks broken in more segments 
(the track reconstruction procedure in KLOE is described in Ref. \cite{offKLOE}). 
We significantly improve the quality of the reconstruction 
requiring the $K^+$ decay
to occur before it reaches the DC sensitive volume, i.e. inside a cylindrical fiducial
volume centered at the IP and with a transverse radius $\rho_{xy}$ close to the
DC inner wall (detector acceptance $\sim$ 26\%).
In this way only the pion tracks are reconstructed, and
we extrapolate only two of them to search for a vertex along the signal kaon path.
No further request on the charge of the particles is applied to maximize the selection efficiency.

\noindent
To extract the number of $K^+\raE \pi^+\pi^-\pi^+(\gamma)$ 
we fit the missing mass spe\-ctrum $m^2_{miss} = E^2_{miss} - (\bold{p}_{K^+} - \bold{p}_1 - \bold{p}_2)^2$
where $\bold{p}_1$ and $\bold{p}_2$ are the momenta of the selected tracks, 
%of the third pion,
with MC-predicted shapes for the signal and the background.
%The selection efficiency is evaluated from MC, and corrections are
%applied to account for data-MC differences on tracking performance.
The branching ratio is given by:
\begin{equation}
BR(K^+ \raE \pi^+\pi^-\pi^+(\gamma)) = \frac{N_{K\raE3\pi}}{N_{tag}}\times\frac{1}{\epsilon_{sel}~ C_{TB} C_{SF} C_{CRV}}
\end{equation}
\noindent
where $N_{K\raE3\pi}$ is the number of signal events, $N_{tag}$ is the number of tagged events and $\epsilon_{sel}$ is
the overall signal selection efficiency, including the detector acceptance and the reconstruction efficiency.
$C_{SF}$ and $C_{CRV}$ are the corrections for the machine background
filter and the cosmic-ray veto. $C_{TB}$ accounts for the tag bias effect.

\subsection{BR measurement with the $K^-_{\mu 2}$ normalization sample}

The normalization sample is given by
$N_{tag}$ = 12065087~ 
%$K^- \raE \mu^- \bar{\nu} (\gamma)$ 
$K^-_{\mu 2}$ tagging events.
The $K^+\raE \pi^+\pi^-\pi^+(\gamma)$ signal selection uses DC information only.

\noindent
Any reconstructed track identified as a $K^+$ 
(and therefore corresponding to a $K^+$ outside the fiducial volume) is rejected.
More specifically we reject tracks with the point of closest approach (PCA) 
to the IP satisfying the conditions
$\sqrt{x^2_{PCA} + y^2_{PCA}} < $ 10 ~cm, and $|z_{PCA}| <$ 20 ~cm,
with the momentum within $70 < p_K < 130$ ~MeV, and with a good matching
with the position and the momentum extrapolated from the tagging kaon.

\noindent
To select the decay vertex $K^+ \raE \pi^+\pi^-\pi^+ (\gamma)$ 
we require at least two reconstructed tracks that have:

\begin{enumerate}
\item  momentum in the kaon rest frame,
          p$^{\ast}_{m_\pi} < $ 190 MeV,
          this cut removes the background from two-body decays;
\item  distance of closest approach (DCA)
          between each extrapolated track and the signal kaon path,
          DCA $<$ 3 cm;
\item  distance of closest approach between the two tracks,
          DCA$_{12} < $ 3 cm;
\item the opening angle between the momenta of the two tracks, $|$cos$(\theta_{12})| <$ 0.9,
          this cut removes the background due to residual kaon broken tracks;
\item the decay vertex is accepted in the fiducial volume, $\rho_{xy} \le$~26~cm. 
\end{enumerate}

\begin{figure}[!thb]
\centering
\includegraphics[width=.6\textwidth]{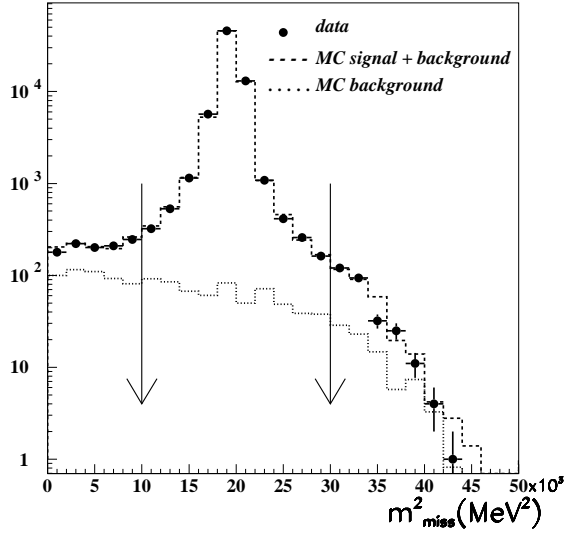} 
\caption[*]{MC (dashed) and data (points) missing mass spectrum of the
selected events. The arrows show the missing mass window 
for signal counting.\label{fig1}}
\end{figure}

\begin{figure}[!thb]
\centering
\includegraphics[width=.6\textwidth]{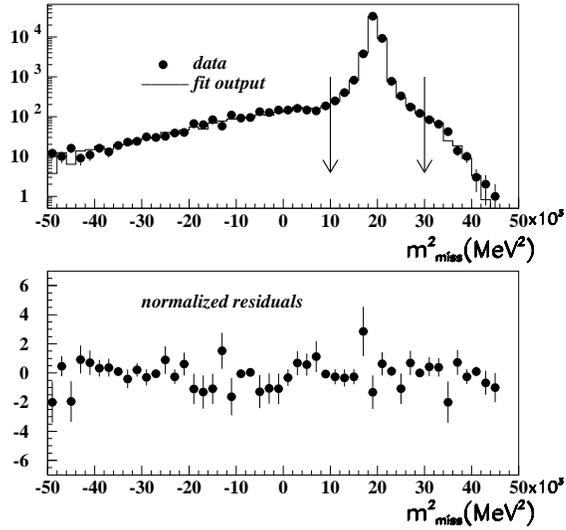}
\caption[*]{Top plot: fit of the missing mass spectrum superimposed
with data points. Bottom plot: 
residuals between the output of the fit 
and data distribution nor\-ma\-li\-zed to their errors. \label{fig2}}
\end{figure}

\noindent
Fig \ref{fig1} shows the comparison between MC and data 
missing mass distributions for the selected $K^+$ decays. 
We count the number of signal events in the
missing mass window $10000 < m^2_{miss} < 30000$ MeV$^2$,
where the signal over background ratio is $S/B \simeq 88$ .
The background composition is given by $K^+$ in 
two-body $\mu^+\nu$ and $\pi^+\pi^0$ $\simeq 0.1\%$, 
semileptonic $\pi^0 e^+\nu$ and $\pi^0\mu^+\nu$ $\simeq 0.5\%$, 
and $\pi^+\pi^0\pi^0 \simeq 0.4\%$ decays. 
These single track events pass the selection criteria because a secondary 
charged track is wrongly reconstructed as two separate tracks.
The top panel of Fig \ref{fig2} shows the result of the fit of the missing mass
distribution compared to data.
The fit gives $N_{K \raE 3\pi}$~=~48032$\pm$286 signal events
(the error accounting for data and MC statistics),
with $\chi^2$/ndf = 44.8/46 (P($\chi^2$) = 0.52). 
The bottom panel of Fig \ref{fig2} shows the fit normalized residuals.

The signal selection efficiency, $\epsilon_{sel}$, is related 
to the track reconstruction efficiency of two 
charged secondaries from $K^+$ decays. 
We evaluate the selection efficiency from MC,
and then we correct it to take into account 
data-MC differences in the track reconstruction.
To this aim we select, both on data and MC, a control sample
of $K^+ \raE \pi^- X$ decays (for signal events X corresponds to $\pi^+ \pi^+$).
The first requirement is the presence of a self-triggering $K^-_{\mu 2}$ 
in the tag hemisphere.
%in order to have homogeneous signal and efficiency samples.
Then the track of the  $\pi^-$ candidate is selected 
with the following requirements:

\begin{enumerate}
\item  the number of neutral clusters 
%         (energy deposits in the EMC not associated to tracks) 
           with an energy $E_{\gamma}\ge$ 30 MeV must be, N$_{clusters} \le$ 1;
\item  the momentum of the selected track in the kaon rest frame must be,
           p$^*_{m_{\pi}} \le$~130~MeV;
\item the distance of closest approach between the extrapolated
          track and the signal kaon path must be, DCA$_{\pi-}$ $<$ 7 cm;
\item the cosine of the opening angle between the momenta of the signal kaon
          and the momenta of the selected track
          must be, cos($\theta_{K\pi})\le$-0.85.
\end{enumerate}

\noindent
The control sample $K^+ \raE \pi^- X$, selected with a background 
contamination of $\simeq$10.7$\%$, is used  
to measure the efficiency corrections as function of the
total transverse momentum $p_X^T$, and of the total longitudinal momentum $p_X^L$ of the $\pi^+\pi^+$ pair
(the average efficiency correction is $\sim$ 0.92).
The selection efficiency, is finally obtained folding the MC selection efficiency 
with the measured corrections: $\epsilon_{sel}$ = 0.0842 $\pm$ 0.0003.

\noindent
The corrections $C_{CRV}$ and $C_{SF}$ have been
measured with data taken without the cosmic-ray veto and the software filter, respectively.
The correction for the tag bias, $C_{TB}$, has been evaluated using MC.
%as the ratio of the branching ratio evaluated after the tag selection and the generated one.
All correction values are reported in  Table \ref{tab1}.

\begin{table}{
\caption{Corrections to BR($K^+\rightarrow \pi^+\pi^+ \pi^-(\gamma)$) measurement.
              The events selected by the two tags have different topologies in the KLOE detector 
              determining different corrections factors.\label{tab1}}
\begin{tabular}{ccc}
\hline\hline
 \bf{Table of corrections}  & $K^-_{\mu 2}$ tags & $K^-_{\pi 2}$ tags  \\
\hline
cosmic ray veto correction $C_{CRV}$  &  1.00125 $\pm$ 0.00002 & 1.00049 $\pm$ 0.00001  \\
software filter correction $C_{SF}$ & 1.0144 $\pm$ 0.0013 & 1.0003 $\pm$ 0.0005 \\
tag bias correction $C_{TB}$  &  0.839 $\pm$ 0.001 & 0.802 $\pm$ 0.002 \\
\hline\hline
\end{tabular}}
\end{table}

\begin{table}[!htbp]{
\small\centering
\caption{Summary table of the fractional statistical uncertainties.\label{tab2}}
\begin{tabular}{ccc}
\hline\hline
\bf {Source of statistical uncertainties} &  $K^-_{\mu 2}$ tags $~(\%$) & $K^-_{\pi 2}$ tags $~(\%$) \\
\hline
signal counting   & 0.45 & 0.70 \\
selection efficiency  & 0.38  & 0.60 \\
tag bias   & 0.11 & 0.18 \\
software filter  & 0.13 & 0.05 \\
cosmic ray veto  & 0.002  & 0.0005 \\
\hline
Total fractional statistical uncertainty  & 0.62 & 0.95  \\
\hline\hline
\end{tabular}}
\end{table}

\noindent
The summary of the fractional statistical uncertainties is reported in Table \ref{tab2}.
The total statistical fractional uncertainty on the branching ratio measurement is $0.62 \%$.

\subsection{BR measurement with the $K^-_{\pi 2}$ normalization sample}

The normalization sample is given by
$N_{tag}$ = 5171239 ~$K^-_{\pi 2}$ tagging events.

\noindent
The signal selection described in sub-section {\it 3.1} is also applied to the sample tagged by $K^-_{\pi 2}$ decays.
The fit to the missing mass spectrum of the selected events gives $N_{K \raE 3\pi}$ = 20,063$\pm$186 signal events 
with $\chi^2$/ndf = 42.9/45 (P($\chi^2$) = 0.56).
The signal over background ratio 
in the missing mass window $10000~<~m^2_{miss} < 30000$ MeV$^2$
is evaluated with MC: $S/B \simeq 84$. 

\noindent
To evaluate the selection efficiency, we used the corrections 
measured with the control sample $K^+ \raE \pi^- X$ tagged by $K^-_{\mu 2}$ events.
The selection efficiency for signal events tagged by $K^-_{\pi 2}$ events, is:
$\epsilon_{sel} = 0.0866 \pm 0.0005$.

\noindent
The summary of the fractional statistical uncertainties is reported in Table \ref{tab2}.
The total statistical fractional uncertainty on the branching ratio measurement 
using the $K^-_{\pi 2}$ tagging sample is $0.95 \%$.

\section{Systematic uncertainties}

The following sources of systematic uncertainties on the branching ratios measured
using both tags, $K^-_{\mu2}$ and $K^-_{\pi2}$, have been considered:

\begin{enumerate}
\item the cuts used to select the signal sample;
\item the fiducial volume;
\item the cuts used to select the control sample $K^+ \raE \pi^- X$;
\item the cuts used to select the tagging samples $K^-_{\mu2}$ and $K^-_{\pi2}$;
\item the charged kaon lifetime.
\end{enumerate}

\begin{table}{
\caption{Summary table of the fractional systematic uncertainties.\label{tab3}}
\begin{tabular}{ccc}
\hline\hline
\bf {Source of systematic uncertainties} &  $K^-_{\mu 2}$ tags $~(\%$)  & $K^-_{\pi 2}$ tags $~(\%$)  \\
\hline
DCA, DCA$_{12}$, cos$(\theta_{12})$ cuts & 0.52  & 0.41 \\
p$^*_{m_{\pi}}$ cut  & 0.08  & 0.11 \\
m$^2_{miss}$ cut &  0.05 & 0.14 \\
fiducial volume  &  0.11 & 0.10 \\
%cuts to select the control sample  $K^+ \raE \pi^- X$  &  0.16 & 0.16  \\
%cuts to select the tagging sample & 0.16 & 0.32 \\
selection efficiency estimate  &  0.16 & 0.16  \\
tag bias & 0.16 & 0.32 \\
$K^{\pm}$ lifetime & 0.12 & 0.12 \\ 
\hline
Total fractional systematic uncertainty  & 0.60 & 0.59  \\
\hline\hline
\end{tabular}}
\end{table}

\noindent
The corresponding systematic uncertainties are listed in Table \ref{tab3}. 

\noindent
The contributions to the systematic error due to points (1), (2), and (3)   
have been evaluated varying the selection cuts. 
%and checking for the BRs stability.
The DCA, DCA$_{12}$ variables and the fiducial volume $\rho_{xy}$ have been
varied within few sigmas, the cuts on cos$(\theta_{12})$,
p$^*_{m_{\pi}}$  and m$^2_{miss}$ have been varied to decrease the $S/B$ 
ratio to $\simeq$ 64. The cuts used to select the control sample $K^+ \raE \pi^- X$ 
have been varied to increase the background contamination up to $\simeq$ 20$\%$.

\noindent
Concerning the selection of the normalization samples (point (4)) 
we have evaluated the effect of a $C_{TB}$ variation 
on the BR measurements.
This has been done modifying 
the selection of the data and MC normalization samples
adding a cut on the opening angle between the
$K^-$ track and the se\-con\-dary track retaining events with cos$(\theta_{Kt}) \ge 0$,
where $t$ is the $\mu^-$($\pi^-$) track in case of the $K^-_{\mu2}$($K^-_{\pi2}$) sample.
Using MC we found that the fra\-ctio\-nal variations of the tag bias corrections are
$\delta C_{TB}/C_{TB} (K^-_{\mu2})= 0.26\%$ and
$\delta C_{TB}/C_{TB} (K^-_{\pi2})= 0.63\%$. Consequently 
the branching ratios measured va\-lues change of 
$\delta BR/BR (K^-_{\mu2}) = 0.32\%$ and $\delta BR/BR (K^-_{\pi2}) = 0.64\%$;
half of these variations have been assigned as conservative values for the  
fractional systematic uncertainties due to the tag bias (see Table \ref{tab3}).

\noindent
The BR($K^+\rightarrow \pi^+\pi^+ \pi^-(\gamma)$) depends
on the charged kaon lifetime  $\tau_{K^{\pm}}$ through the detector acceptance,
that is evaluated with MC simulation (point (5)).
The systematic effect has been obtained
varying $\tau_{K^{\pm}}$ within the uncertainty
of the KLOE result $\tau_{K^{\pm}} = 12.347 \pm 0.030$ ns \cite{Kpmlife}.
This has been done re-weighting the
MC events with a hit-or-miss procedure, both for the signal and the control sample
selection procedures.
%Consequently the BR depends on  $\tau_{K^{\pm}}$ as :
%\begin{equation}
%BR^{\tau_{K^{\pm}}}/BR^{(0)} = 1 - 0.0412~{\rm ns}^{-1}(\tau_{K^{\pm}} - \tau^{(0)}_{K^{\pm}})
%\end{equation}
%\noindent
%where $\tau^{(0)}_{K^{\pm}}$ = 12.38 ns is the current PDG fit value \cite{PDG}. 
The corresponding sistematic errors are listed in Table \ref{tab3}.

\noindent
The analysis is fully inclusive of radiative decays. 
Only the efficiency evaluation could  be affected by a systematic uncertanty 
due to the cut  N$_{clusters} \le $ 1  (see sub-section ${\it 3.1}$).
We have used PHOTOS \cite{photos} to evaluate such an effect and we obtained a 
negligible contribution, being $O(10^{-6})$ the fraction of  decays removed 
by the cut N$_{clusters} \le $ 1. 

\noindent
The fraction of $K^+$ undergoing nuclear interactions is
negligible, $\sim$~10$^{-5}$, as evaluated using
the MC simulation, based on data available in literature \cite{Knuclear}. 
Therefore the related systematic uncertainty is negligible.

\noindent
Furthermore we have checked on  two independent
sub-samples of about 88 pb$^{-1}$ and 86 pb$^{-1}$ that the efficiency corrections and the
BR evaluations are not correlated.

\noindent
Finally the stability of the measurements with respect to different
data taking periods and conditions has been checked.

\section{Results}

With a sample of $K^- \raE \mu^- \bar{\nu} (\gamma)$ tagging events $N_{tag}$ = 12065087 we found 
$N_{K \raE 3\pi} = 48032 \pm 286$ signal events.  Using equation 1, we obtain the branching ratio:

\begin{equation}
BR(K^+ \raE \pi^+\pi^-\pi^+(\gamma))|_{Tag K_{\mu2}} = 0.05552 \pm 0.00034_{stat} \pm 0.00034_{syst}.
\end{equation}

\noindent
With a sample of $K^- \raE \pi^- \pi^0 (\gamma)$ tagging events $N_{tag}$ = 5171239 we found 
$N_{K \raE 3\pi} = 20063 \pm 186$ signal events, corresponding to:

\begin{equation}
BR(K^+ \raE \pi^+\pi^-\pi^+(\gamma))|_{Tag K_{\pi2}}  = 0.05587 \pm 0.00053_{stat} \pm 0.00033_{syst}.
\end{equation}

Averaging these two results, accounting for correlations, we obtain:

\begin{equation}
BR(K^+ \raE \pi^+\pi^-\pi^+(\gamma)) = 0.05565 \pm 0.00031_{stat} \pm 0.00025_{syst}.
\end{equation}

\noindent
This absolute branching ratio measurement is fully inclusive of
final-state radiation and has a 0.72$\%$ accuracy, 
a factor $\simeq$ 5 better  with respect to the previous measurement \cite{chiang}.

\begin{table}[!h!b!t]{
\caption{Results of the fit: $K^{\pm}$ BRs and correlation coefficients.\label{tab4}}
\begin{tabular}{llccccccc}
\hline\hline
Parameter & Value & \multicolumn{6}{c}{Correlation coefficients} \\
\hline
BR($K^{\pm}_{\mu2}$) & 0.6372(11)               &          &       &    &    &    &   \\
BR($K^{\pm}_{\pi2}$) &  0.2070(9)                  &  0.55 &       &    &    &    &   \\
BR($\pi^{\pm} \pi^- \pi^+$) &  0.0558(4)      & -0.23 & -0.05 &    &    &    &   \\
BR($K^{\pm}_{e3}$) &  0.0498(5)                     &   0.42 &  -0.15 &  0.06 &    &    &   \\
BR($K^{\pm}_{\mu3}$) & 0.0324(4)                 &  -0.39 &  0.14 & -0.05 & -0.58 &    &   \\
BR($\pi^{\pm} \pi^0 \pi^0$)  & 0.01764(25)  &  -0.13 &  0.05 &  -0.02 & 0.04 & -0.04   &   \\
$\tau_{K^{\pm}}$ (ns) &   12.344(29)              &   0.20 &   0.19 & -0.14 & 0.05 & -0.04&  0.02 \\
\hline\hline
\end{tabular}}
\end{table}

\noindent
We fit the six largest $K^{\pm}$ BRs and the lifetime $\tau_K^{\pm}$ using the KLOE measurements of
$\tau_{K^{\pm}}$ \cite{Kpmlife},
BR($K^+_{\mu2}$) \cite{Kmunu}, BR($K^+_{\pi2}$) \cite{KLOE_K3piFit}, 
BR($K^+ \raE \pi^+\pi^-\pi^+(\gamma)$) (eq. 4), BR($K^{\pm}_{l3}$) \cite{KpmSemil}, and
BR($K^{\pm} \raE \pi^{\pm}\pi^0\pi^0$)  \cite{Kpmpch2p0}, with their
dependence on $\tau_K^{\pm}$, and 
imposing the constraint $\sum$BR$(K^{\pm} \raE f)$~=~1.
%the sum of the BRs constrained to unity.
The fit results, with $\chi^2$/ndf = 0.24/1 (CL = 0.63),
show a coherent set of measurements (see Table \ref{tab4}).

\noindent
The NA48 experiment observed in the $\pi^0\pi^0$ invariant mass distribution
a cusp-like anomaly at
$M_{00} = 2m_{\pi^+}$ \cite{NA48-1}, which has been interpreted
as mainly due to the final state charge-exchange reaction 
$\pi^+\pi^- \rightarrow \pi^0\pi^0$ in 
$K^{\pm} \rightarrow \pi^{\pm}\pi^+\pi^-$ decay \cite{Budini},  \cite{Cabibbo}.
The fit to the $M_{00}^2$ distribution \cite{NA48-2} with two different
models \cite{Cabibbo-Isidori}  and \cite{Berna1} \cite{Berna2} determines 
 $a_0-a_2$, the difference
between the S-wave $\pi\pi$ scattering lengths in the isospin $I$=0 and $I$=2 states. 
In this calculation the main source of uncertainty is the ratio of the weak
amplitudes of $K^{\pm} \raE \pi^{\pm}\pi^-\pi^+$ and $K^{\pm} \raE \pi^{\pm}\pi^0\pi^0$ decay,
that is obtained from the ratio $R$ of the branching ratio values.
%Using the BRs in Table \ref{tab4}
Using the BR($\pi^{\pm} \pi^- \pi^+$), BR($\pi^{\pm} \pi^0 \pi^0$) 
and their correlation shown in Table \ref{tab4}
we evaluate $R = 3.161 \pm 0.049$, in agreement with the value 
$R = 3.175 \pm 0.050$ obtained by NA48 \cite{NA48-2} with BRs from the
PDG fit \cite{PDG}.

\section{Conclusions}

We have measured the absolute branching ratio of the $K^+ \raE \pi^+\pi^-\pi^+(\gamma)$ decay,
inclusive of final-state radiation, using two indipendent normalization samples from 
$K^-_{\mu 2}$ and $K^-_{\pi 2}$ tags:

\[
BR(K^+ \raE \pi^+\pi^-\pi^+(\gamma)) = 0.05565 \pm 0.00031_{stat} \pm 0.00025_{syst}
\]

\noindent
with an overall accuracy of  0.72$\%$.
This measurement completes the KLOE program of
precision measurements of the dominant kaon branching ratios.

%\title{Acknowledgements (19 October 2012)}
%\author{The KLOE-2 collaboration}
%\date{}                                           % Activate to display a given date or no date
%\begin{document}
%\maketitle
\section{Acknowledgements}
%\subsection{}
We warmly thank our former KLOE colleagues for the access to the data 
collected during the KLOE data taking campaign.
We thank the DA$\Phi$NE team for their efforts in maintaining low background running conditions 
and their collaboration during all data taking. We want to thank our technical staff: 
G.F. Fortugno and F. Sborzacchi for their dedication in ensuring efficient operation of the KLOE computing facilities; 
M. Anelli for his continuous attention to the gas system and detector safety; 
A. Balla, M. Gatta, G. Corradi and G. Papalino for electronics maintenance; 
M. Santoni, G. Paoluzzi and R. Rosellini for general detector support; 
C. Piscitelli for his help during major maintenance periods. 
This work was supported in part by the EU Integrated Infrastructure Initiative Hadron Physics Project under contract number RII3-CT- 2004-506078; by the European Commission under the 7th Framework Programme through the `Research Infrastructures' action of the `Capacities' Programme, Call: FP7-INFRASTRUCTURES-2008-1, Grant Agreement No. 227431; by the Polish National Science Centre through the Grants No. 
%0469/B/H03/2009/37, 
%0309/B/H03/2011/40, 
DEC-2011/03/N/ST2/02641, 
2011/01/D/ST2/00748,\\
2011/03/N/ST2/02652,
2013/08/M/ST2/00323,
and by the Foundation for Polish Science through the MPD programme and the project HOMING PLUS BIS/2011-4/3.





\end{document}